\documentclass[prl,twocolumn,superscriptaddress,showpacs]{revtex4}
\usepackage{graphicx}
\begin{document}
\title{Fluctuation-Dissipation Theorem for Metastable Systems}

\author{G.~B\'aez}
\email{baez@fis.unam.mx}
\affiliation{Centro de Ciencias F\'{\i}sicas UNAM, A.P 48-3, 62210, Cuernavaca, Morelos, M\'exico}
\affiliation{Laboratorio de Sistemas Din\'amicos, Departamento de Ciencias B\'asicas,
Universidad Aut\'onoma Metropolitana-Azcapotzalco, A.P. 21-267, Coyoac\'an 04000, M\'exico D.F., M\'exico}
\author{H.~Larralde}
\affiliation{Centro de Ciencias F\'{\i}sicas UNAM, A.P 48-3, 62210, Cuernavaca, Morelos, M\'exico}
\author{F.~Leyvraz}
\affiliation{Centro de Ciencias F\'{\i}sicas UNAM, A.P 48-3, 62210, Cuernavaca, Morelos, M\'exico}
\author{R.~A.~\surname{M\'endez-S\'anchez}}
\affiliation{Centro de Ciencias F\'{\i}sicas UNAM, A.P 48-3, 62210, Cuernavaca, Morelos, M\'exico}

\begin{abstract}
We show that an appropriately defined fluctuation-dissipation theorem,
connecting generalized susceptibilities and time correlation
functions, is valid for times shorter than the nucleation time of the
metastable state of Markovian systems satisfying detailed balance. This is done by assuming that such systems can be
described by a superposition of the
ground and first excited states of the master equation.
We corroborate our results numerically for the
metastable states of a two-dimensional Ising model.
\end{abstract}
\maketitle

There have been many efforts to extend the concepts and methods used to
describe systems in equilibrium to systems which are not in equilibrium but
are either stationary, or evolve very slowly
\cite{Sciortino&Tartaglia,Ismailov&Myerson}.  A particular class of ``slowly
evolving'' out of equilibrium systems are those which are in a metastable
state and, due to their ubiquity, their characterization is of especial
interest. Usually it is thought that the macroscopic properties of a metastable
system can be treated as if it was in equilibrium. In particular,
even relations such as the fluctuation dissipation theorem (FDT)
\cite{Kubo57,Kubo59,Kubo66} are generally assumed to be valid for
systems in a metastable state.  However, metastable systems are
actually far from equilibrium and there is no reason to expect the
validity of this theorem for such systems, even if their evolution is
very slow. Indeed, the FDT does not apply to systems such as
finite-range spin glasses, domain growth processes, structural
glasses, among others (See Ref.~\cite{Grigera&Israeloff} and
references therein). In this letter we use a dynamical approach to
show why it is justified to apply results from equilibrium to
metastable states for the case of Markovian stochastic dynamics and we
derive the FDT for these systems from the microscopic dynamics.

Since the phenomenology of metastable states has been assumed
to be similar to that of equilibrium systems, most of the efforts have focused
on understanding the mechanisms by which a system decays from the metastable
state to equilibrium by nucleation processes (growing of a second phase)
\cite{Langer,nucleacion1,nucleacion2,campom1,Lebowitz}.
However, a theory for the description of metastable states {\it per se} is still lacking
\cite{Penrose-Lebowitz}. This is partly because the phenomenon of
metastability is a relative and rather complicated concept
\cite{Debenedetti}.  Penrose and Lebowitz \cite{Lebowitz},
made a detailed characterization of the principal properties observed
in the behavior of systems in a metastable state, which can be
summarized as follows: In a metastable state, a system behaves similarly
to a hypothetical pure
thermodynamic phase, although the intensive parameters have
values such that the equilibrium state would consist of a different
phase or a coexistence of different phases.  When the system is
isolated, the metastable state remains for a very long time. The
response to small and slow perturbations leads to small and reversible
changes in the systems. For large or rapid changes, the system may
escape irreversibly from the metastable state. Beyond qualitative
characterizations, there is not a clear and general definition of
metastability \cite{Neves&Schonmann}.
In this work we use a definition of a
metastable state similar to that introduced by Davies \cite{Da1,Da2} for Markovian systems
satisfying detailed balance, in terms of the eigenvectors of the
corresponding time independent master equation \cite{Glauber}.
Using this definition and the Kubo formalism for linear response theory
\cite{Kubo57,Kubo59,Kubo66}, we obtain a metastable fluctuation-dissipation
theorem valid for times short compared with the nucleation time of
the system.

In the following we limit ourselves to the dynamics of Markovian systems with a
finite (possible large) number of states. Those can be described by a master equation
to which we associate an operator $\hat L_0$
\begin{equation}
{ \dot P}(t)=\hat L_0  P(t), \label{operator}
\end{equation}
If the system is characterized by a set of discrete random variables
$\vec \sigma$, then $P(t)$ is a vector of components $p(\vec\sigma,t)$
which correspond to the probability that the system is in the
state specified by $\vec \sigma $ at time $t$.
When  Eq.~(\ref{operator}) is solved by separation of variables,
we obtain the time independent master equation
\begin{equation}
\hat L_0 \psi_j  =  -\Omega_j \psi_j ,
\label{timeindependent}
\end{equation}
where $\Omega_j$ and $\psi_j$ are the eigenvalues and eigenvectors of
$\hat L_0$, respectively. Due to conservation of probability
there exists a stationary
solution $\psi_0(\vec\sigma)$ associated to the eigenvalue
$\Omega_0=0$, namely $ \psi_0(\vec\sigma)=p^e(\vec\sigma)$.
Here $p^e$ is the Boltzmann probability distribution of the system
in the equilibrium state, which we assume to be the only stationary
solution of the master equation.
If detailed balance holds
then $\hat L_0$ is self-adjoint with respect to the following internal product
\begin{equation}
(R,Q)=\sum_{\vec\sigma}\frac{R(\vec\sigma) Q(\vec\sigma)}
{\psi_0(\vec\sigma)},
\label{productopunto}
\end{equation}
with $R$ and $Q$ two arbitrary functions with finite norm\cite{Van Kampen}.
For $R=\psi_0$ and $Q=\psi_j$, Eq.~(\ref{productopunto}) implies
$\sum_{\vec\sigma}\psi_j(\vec\sigma)= \delta_{0,j}$, i. e. $\psi_0(\vec\sigma)$
is normalized and $\psi_{j\ne0}(\vec\sigma)$ sum to zero.

We assume now that it is possible to choose the parameters of the system in such a way
that one of the eigenvalues of $\hat L_0$, labeled by $-\Omega_1$,
corresponds to a decay that is much slower than the observational times. This is
$0 < \Omega_1\ll 1\ll \Omega_j$, for $j\ge 2$, in appropriate units. This assumption means that
we neglect the case of having several different metastable states. The extension to finitely many
metastable states is straightforward, but not the one to systems with a divergent number of
metastable states (glasses, spin glasses, etc.).

Let us now prepare the system in any configuration $\vec\sigma'$.
Since the set of eigenfunctions is complete (as follows from the self-adjointness of
$\hat L_0$), we can represent the corresponding
probability distribution as
\begin{equation}
\delta_{\vec\sigma,\vec\sigma'}=\psi_0(\vec\sigma)+\sum_{j=1}^{\infty}
\frac{\psi_j(\vec\sigma')\psi_j(\vec\sigma)}{\psi_0(\vec\sigma')},
\label{delta}
\end{equation}
and hence, for $1/\Omega_2  \ll t \ll1/\Omega_1 $, one gets a nearly stationary state
(which essentially does not vary in time for $t\ll 1/\Omega_1$)
\begin{equation}
e^{\hat L_0 t} \delta_{\vec\sigma,\vec\sigma'}\approx \psi_0(\vec\sigma)+
G(\vec\sigma') \psi_1(\vec\sigma),
\label{initialcondition}
\end{equation}
where $G(\vec\sigma')=\psi_1(\vec\sigma')/\psi_0(\vec\sigma')$. That
the RHS of Eq.~(\ref{initialcondition}) is a probability distribution follows from
the norm and positivity preserving properties of $\exp(\hat L_0 t)$.
When $G(\vec\sigma')\ll 1$, the state given by
Eq.~(\ref{initialcondition}) is the equilibrium state, since the
second term in the RHS is negligible. On the other hand, when
$G(\vec\sigma')\gg1$ the state is sharply localized in a zone of configurations
$\{\vec\sigma\}_m$ (hereafter, the metastable {\it zone}) and
very small outside this zone. Then, this quasistatic probability distribution represents the metastable state. 
Notice that in this situation $G$ is
independent of $\vec\sigma'$ since any configuration prepared within the
metastable zone is expected, on physical grounds, to evolve into the same intermediate
metastable state $p^m(\vec\sigma)$.  The case $G(\vec\sigma')\sim1$, leads to configurations which have comparable
probabilities of evolving into either the equilibrium or metastable
state. We assume that this set of ``saddle points''
is negligible compared to the sets of both, equilibrium and metastable configurations.


Now consider a system which can be prepared in a metastable initial state
described by
\begin{equation}
 p^m(\vec\sigma)=\psi_0(\vec\sigma)+G \psi_1(\vec\sigma); \qquad G\gg1.
\label{metastable}
\end{equation}
As $p^m(\vec\sigma)$  is
negligible outside the metastable zone, we approximate  $p^m(\vec\sigma)$  as
\begin{equation}
p^m(\vec\sigma)\approx
\left\{\begin{array}{ll}
\psi_0(\vec\sigma)+G \psi_1(\vec\sigma)
&\mbox{for $\vec\sigma\in\left\{\vec\sigma\right\}_m$}\\
0 &\mbox{for
$\vec\sigma\not\in\left\{\vec\sigma\right\}_m$}
\end{array}
\right. ,
\label{approximate}
\end{equation}
with $G\gg1$.
The last equation coincides with the definition of the metastable state given by
Davies \cite{Da1,Da2}, where the reader can find greater detail.

Using the definition of $G$ and the fact that it is constant within the metastable
zone one gets $\psi_1(\vec\sigma)=G \psi_0(\vec\sigma)$ for
$\vec\sigma\in\left\{\vec\sigma\right\}_m$. On the other hand, using that
$p^m(\vec\sigma)=0$ outside the metastable zone, we get
$\psi_1(\vec\sigma)=-1/G \psi_0(\vec\sigma)$ for
$\vec\sigma\notin\left\{\vec\sigma\right\}_m$
Thus, the first excited state $\psi_1(\vec\sigma)$, and hence, the
metastable state, is specified in terms of the equilibrium
distribution for (almost) all configurations of the system
since it is locally proportional to the Boltzmann distribution
in both, metastable and not metastable zones.
The proportionality coefficients are given by $G$ and $-1/G$, respectively.

This simple picture of metastability allows us to go beyond the description of the distributions  characterizing
the metastable states. In particular, we now derive a FDT through
linear response theory for these states. We now  consider the perturbed
master equation:
\begin{equation}
\dot P + h \dot{P_1}= \left(\hat L_0+ h  e^{i\omega t}
\hat L_1 \right)
\left(P+h P_1 \right),
\label{perturba}
\end{equation}
where $\hat L_1$ is the perturbative term generated by an oscillatory
external field and
$P$ is the probability distribution in the absence of
the perturbing external field, whose evolution is described by Eq.~(\ref{operator}).

Now, if the system is initially in its metastable state ---described by
~Eq.~(\ref{metastable})--- after some algebra, we obtain the following general expression
for the changes in the probability distribution $P_1(\vec\sigma,t)$, to first order in $h$,
\begin{eqnarray}
&&P_1 (\vec\sigma,t) = h\int_0^t e^{\hat L_0(t-t')}
\hat L_1 \left[ \psi_0(\vec\sigma)+
G\psi_1(\vec\sigma) \right] e^{i\omega t'}  dt'  \nonumber \\
&&  + h\int_0^t e^{\hat L_0(t-t')}\hat L_1 \  G\psi_1(\vec\sigma)\
e^{i\omega t'} (e^{-\Omega_1 t'}-1) dt'
\label {deltaPt}
\end{eqnarray}
This expression is exact for all times. Since $\hat L_1$, $\psi_0$ and $\psi_1$
are independent of $t'$ the integrations are trivial.
Expanding $P_1 (\vec\sigma,t)$ in the basis of eigenvectors of $\hat L_0$,
the equilibrium case is recovered for $t\rightarrow \infty$. For times
$t\ll 1/\Omega_1$, the second integral vanishes, and the first
integral yields the total change of the probability distribution
starting from the metastable initial condition.

Introduce for any $\rho$ the following notation
\begin{equation}
\langle B(t)\rangle_{\rho} =
\sum_{\vec\sigma} B(\vec\sigma)\rho(\vec\sigma,t).
\label{average}
\end{equation}
We can calculate the changes of the average value of
any physical quantity $B(\vec\sigma)$ as $\langle B(t)\rangle_{P_1}$ where $P_1$ is taken
from Eq.~(\ref{deltaPt}).

By taking the corresponding Laplace-Fourier transform
of $\langle B(t)\rangle_{P_1} $,
we can define \cite{definition} the
metastable susceptibility of the system as
\begin{eqnarray}
\chi(s)&&=\sum_{\vec\sigma} B(\vec\sigma)\sum_j
\psi_j(\vec\sigma) \left\{ \frac{ \left(\psi_j , \hat L_1
(\psi_0+G\psi_1) \right) }{\left(s+\Omega_j \right) } \right .  \nonumber \\ &&-
\left . \frac{ \Omega_1 \left( \psi_j , \hat L_1\  G\psi_1 \right)}{(s+
\Omega_j)(s+\Omega_1-i\omega)} \right\}.
\label{susceptibility}
\end{eqnarray}

Then, in linear approximation, the metastable susceptibility consists of
two terms. The first term is similar to the equilibrium case but this time
the initial condition is the probability distribution of the
metastable state. The second is a memory term corresponding to a
convolution with the external field.


We now define $\psi_j(h)$ as the eigenfunctions of the operator
$L_0+h\hat L_1$. To first order in $h$ we have the
following relation connecting $\hat L_1\psi_i(0)$ with the
derivatives of $\psi_1(h)$ with respect to $h$:
\begin{equation}
\hat L_1\psi_i(0)=-\left.\hat L_0\frac{\partial\psi_i}{\partial
h}\right|_{{h=0}}-\left.\Omega_i(0)\frac{\partial\psi_i}{\partial
h}\right|_{{h=0}}
-\left.\frac{\partial\Omega_i}{\partial h}\right|_{h=0}\psi_i(0)
\label{variation}
\end{equation}

We substitute Eq.~(\ref{variation}) in
the scalar products of Eq.~(\ref{susceptibility}) and use
the appropriate
proportionality between $\psi_1(\vec\sigma)$ and $\psi_0(\vec\sigma)$.
We then split the sum over
$\vec \sigma$ in a sum over the stable zone and one over the metastable zone
(See Eq.~\ref{approximate}),
noting that if the system has
a metastable state as defined above, then the higher
excited states satisfy $\sum_{\vec\sigma_m}
\psi_j(\vec\sigma_m)= \sum_{\vec\sigma \not\in\{\vec\sigma_m\}}
\psi_j(\vec\sigma)\approx 0$,\- for $j\ge 2$.  This must be the case as
$\sum_{\vec\sigma\in\{\vec\sigma_m\}} P(\vec\sigma,t)=1$ for
times shorter than the nucleation time $1/\Omega_1$.

We now define $E(\vec\sigma)$ as the energy of the system appearing in the Boltzmann distribution and
introduce, for each probability distribution $\rho(\vec\sigma,t)$, the following dynamical correlation
\begin{equation}
\left\langle\dot B(t) \frac{\partial E}{\partial h}\right\rangle_{\rho}=
\sum_{\vec\sigma}\frac{\partial E (\vec\sigma)}{\partial h}\rho(\vec\sigma,t)\sum_{\vec\sigma'}B(\vec\sigma')
\dot p(\vec\sigma', t' | \vec\sigma,t ),
\end{equation}
where $p(\vec\sigma',t' | \vec\sigma,t)$ is the conditional probability that the configuration $\vec\sigma'$ occurs at time $t'$
given that it was in $\vec\sigma$ at time $t$. After several pages of algebra one then gets for the
metastable susceptibility
\begin{widetext}
\vspace*{-7mm}\begin{equation}
\chi(s)=\beta{\cal L}\mathbf{\Bigg[}
\left \langle\dot B(t) \frac{\partial E}{\partial h}\right\rangle_{p^m} 
+\Omega_1 \Delta\left\langle  B(t) \frac{\partial}{\partial h} (E-kT\ln\Omega_1) \right\rangle 
+   \Omega_1 \int_0^t d\tau\, e^{iw(t-\tau)}\Delta\left\langle \dot B(t) \frac{\partial E}{\partial h} \right\rangle
+o(\Omega_1)
\mathbf{\Bigg]}
\label{ximeta2}
\end{equation}
\vspace*{-2mm}
\end{widetext}
\vspace*{-10mm}where $\beta=1/kT$ and
$\Delta\langle A \rangle=\langle A \rangle_{p^m}-\langle A \rangle_{p^e}$.

To obtain the metastable fluctuation-dissipation theorem it is enough to show that
the second and third terms in Eq.~(\ref{ximeta2}) vanish as the coexistence
curve is approached (first order correction in $\Omega_1$).
Since $B(t)$ and $\dot B(t)$ remain bounded, the terms related to them
are negligible because $\Omega_1 \ll 1$.
Indeed, the principal correction is given by the
term of the order of $\partial \ln \Omega_1/\partial h$.
Since $\Omega_1$ is the nucleation rate\cite{Langer}, it is given roughly by
$\exp(-\beta W)$ where
$W$ is the nucleation barrier. Usually $W$ is roughly
$R^{d-1}_c$, where $R_c$ is the critical droplet radius and diverges algebraically as the
supersaturation $h_0$ goes to zero. Here the total strength of the external field is given
by the fixed initial field $h_0$ plus the perturbation $h$. ($h_0$ in the Ising model is the external magnetic field).
From this follows that $\partial \Omega_1/\partial h$ also
diverges algebraically in $h$, whereas $\Omega_1$ goes to zero as a stretched exponential.
{\em Thus we have established the central result of this work,
a fluctuation dissipation theorem for the metastable states}.

\begin{figure}
\includegraphics[width=0.9\columnwidth,angle=0]{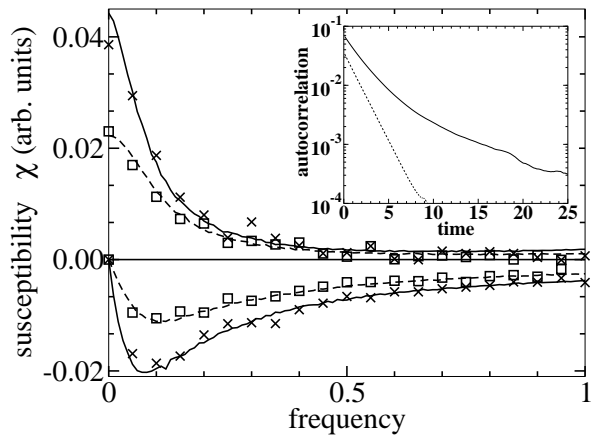}
\caption{Magnetic susceptibility for a two dimensional Ising model with $10^4$ spins.
The temperature, $T=1.52$, external magnetic field $h_0=0.1$ and coupling $J=1$.
As usual, the real part is the
positive quantity and the imaginary part is the negative.
The crosses correspond to the metastable computation by the perturbative
method and the empty squares to the equilibrium one.
The solid and dashed lines were obtained by transforming the metastable and equilibrium
autocorrelations of the magnetization. The inset shows the autocorrelation of the magnetization.
The solid and dashed lines correspond to the metastable  and equilibrium
correlations respectively.}
\label{figure2}
\end{figure}

In order to check the FDT for metastable states, we will show that the susceptibility obtained by
perturbing the metastable state of a two dimensional Ising model with a magnetic field of fixed frequency,
agrees with that obtained by
taking the Fourier-Laplace transform of the correlations of the fluctuations in the metastable state,
given by
the first term in the r.h.s. of Eq.~(\ref{ximeta2}). Our programs where proved checking the
well known FDT in equilibrium (See Fig.~\ref{figure2}). We obtained the
autocorrelation of the magnetization by a  Monte Carlo
simulation for a two dimensional Ising model evolving by Glauber dynamics
\cite{Glauber}
on a square lattice with periodic boundary conditions.  The set of
external parameters (temperature $T$, coupling $J$ and static magnetic field $h_0$) was chosen
to give long-lived metastable states
($\sim 10^4$ Monte Carlo time steps per spin) when starting with all spins
opposite to the magnetic field. As suggested in Ref.~\cite{NotesinComp,Ma}, we used
$T=\frac 23 T_c J$, where $T_c$ is the critical temperature and $J$ is the coupling
between nearest neighbor spins. The external magnetic field
$h_0$ was chosen by trial.
For all calculations we allowed the system to evolve until a long-lived state
opposite to the field was reached.
Then we computed the average of the fluctuations of the magnetization
with respect to the  metastable equilibrium value, over a set of
100 realizations.

In Fig.~\ref{figure2} we show the magnetic susceptibilities as a
function of the frequency. The empty squares correspond to the classical magnetic
susceptibility for the equilibrium state, obtained by the definition
of linear response theory. The crosses correspond to the same
quantity for the metastable state. Both quantities were averaged over
10 realizations. The dashed lines correspond to the
equilibrium magnetic susceptibility computed using Laplace-Fourier
transformation of the corresponding autocorrelations (per spin)
of the magnetization given in the inset (dashed line).  Finally the solid
lines correspond to the magnetic susceptibility obtained by
transforming the appropriate autocorrelation function in the
metastable state (continuous line in the inset). We find the same
good level of agreement for the equilibrium and metastable cases.
It is interesting to observe that the behavior of the correlations are
very different in both cases.

In summary, by using a formal definition of a metastable state
for the case of Markovian systems with a finite number of states
we have shown that the FDT indeed holds for times shorter than the nucleation
time.
We also evaluated the size of the leading corrections.
Since many systems have
Markovian dynamics on sufficiently long time scales, this result
has quite a broad range of applicability.
A crucial hypothesis was the
existence of one single low-lying excited state of the
operator $\hat L_0$. This means that nucleation is the slowest physical
process, a condition often satisfied in practice.
Detailed calculations
and numerical simulations will be given elsewhere \cite{Baezetal}.

We thank A. L. Salas-Brito for reading the manuscript.
G. B\'aez acknowledges the support of the CONACyT (Project 32173-E)
and DGAPA-UNAM (Project IN-112200).

\bibliographystyle{prsty}
\bibliography{thesis,paperdef,paper,newpaper,book}

\begin{thebibliography}{00}
\bibitem{definition}Notice that this definition uses Fourier-Laplace instead of Fourier transform
since the metastable states decay for $t\rightarrow \infty$.
\end{thebibliography}

\end{document}